\documentclass[twocolumn,prl,notitlepage,preprintnumbers,superscriptaddress,amsmath,amssymb]{revtex4-2}

\usepackage{graphicx} 
\usepackage[hidelinks]{hyperref}

\graphicspath{{./Figures/Paper/}}

\usepackage{xcolor}
\usepackage[normalem]{ulem}

\newcommand{\ket}[1]{\lvert #1 \rangle}

\newcommand{\avg}[1]{\langle #1 \rangle}

\newcommand{\fLC}{f_{\text{LC}}}

\newcommand{\trev}{t_{\rm rev}}
\newcommand{\tGate}{t_{\rm gate}}

\newcommand{\HLCJ}{H_{\rm LCJ}}
\newcommand{\Heff}{H_{\rm eff}}
\newcommand{\Veff}{V_{\rm eff}}
\newcommand{\NDD}{N_{\rm DD}}

\newcommand{\xiphi}{\xi_{\varphi}}
\newcommand{\gammaphi}{\gamma_{\varphi}}

\renewcommand{\mod}{\,\,\text{mod}\,\,}

\DeclareMathOperator{\sgn}{sgn}

\begin{document}

\title{Exponentially robust non-Clifford gate in a driven-dissipative circuit}
\author{Liam O'Brien}
\affiliation{
Department of Physics and Institute for Quantum Information and Matter, California Institute of Technology, Pasadena, CA 91125, USA}
\author{Gil Refael}
\affiliation{
Department of Physics and Institute for Quantum Information and Matter, California Institute of Technology, Pasadena, CA 91125, USA}
\affiliation{AWS Center for Quantum Computing, Pasadena, CA, 91125, USA}
\author{
Frederik Nathan}
\affiliation{Center for Quantum Devices and NNF Quantum Computing Programme, Niels Bohr Institute,  University of Copenhagen, 2100 Copenhagen, Denmark}
\affiliation{
Department of Physics and Institute for Quantum Information and Matter, California Institute of Technology, Pasadena, CA 91125, USA}

\begin{abstract}
    Recent work (Nathan \textit{et al}, \textit{arXiv:2405.05671}) proposed an architecture for a dissipatively stabilized GKP qubit, and protocols for protected Clifford gates. Here we propose a protocol for a protected  non-Clifford $\sqrt{T}$ gate at the physical qubit level, based on the inclusion of a quartic flux potential generated by ancillary Josephson junctions. We show that such a gate is topologically robust with exponentially suppressed infidelity from control or device imperfections, and operates on microsecond timescales for GHz resonators. We analyze the resilience of the protocol to noise, imperfect control, and imperfect targeting of circuit parameters.   
\end{abstract}

\maketitle




To achieve quantum advantage, a quantum computer must support non-Clifford gates, which are intrinsically hard to implement fault tolerantly~\cite{Gottesman-GK-theorem,Aaronson-Gottesman-GK,Kliuchnikov-Maslov-Mosca-Single-Qubit-Clifford-with-T,Forest-Gosset-Kliuchnikov-McKinnon-Single-Qubit-general,Kitaev-Universal-Computation,Dawson-Nielsen-SK-algo,EastinKnill}. Current approaches rely on  \textit{magic state distillation}~\cite{Bravyi-Kitaev-magic-states}, which, despite algorithmic improvements~\cite{Fowler-Gidney-lattice-surgery,Bravyi_2012,Bravyi-Haah-distillation-triorthogonal,Fowler-Devitt-distillation-surface-code,Meier-Eastin-Knill-distillation-four-qubit,Jones-distillation-multilevel,Duclos-Svore-distillation-arbitrary-nonstabilizer,Duclos-Pulin-distillation-complex-gates,Campbell-Howard-distillation-and-gate-synthesis,OGorman-Campbell-distillation-block-code-realistic,Haah-Hastings-mutliple-distillations,Campbell-Howard-distillation-nonPauli-parity,Gidney-Fowler-distillation-CCZ-and-T,Litinski-distillation-not-as-costly},  involves significant overhead. As a result, they are a key anticipated bottleneck of useful quantum computation \cite{Babbush-Gidney-Berry-Wiebe-McClean-Paler-Fowler-Neven-electronic-spectra,Calderbank-Rains-Shor-Sloane-Clifford,Gottesman-Fault-Tolerant-QC}, motivating ongoing  search for alternative 
approaches \cite{Jones-Brooks-Harrington-gauge-color-codes,Bravyi-Cross-doubled-color-codes,Jochym-Bartlett-Stacked-codes,Bombin-3d-color-codes,Chamberland-Cross-flag-qubits,Brown-dynamic-measurements-surface-code,Low-Kliuchnikov-Schaeffer-trade-T-gates}. 
Another current challenge of quantum computing is achieving  high qubit quality.
Recently, several groups discovered a new approach to this goal~\cite{LachanceQuirion-2023,Lev-Arcady-Phillipe-GKP-Proposal,gkp-paper,Max-Frederik-GKP} via self-correcting Gottesman-Kitaev-Preskill (GKP) qubits \cite{GKP} that leverage dissipation to  correct bit flip \textit{and} dephasing errors. The qubits moreover support native single-qubit Clifford gates, that are {\it topologically robust}---in the sense that errors remain exponentially small with finite control noise and mistargeting of circuit parameters.   Some of the self-correcting GKP qubits may be within reach of near term devices~\cite{campagne-ibarcq-gkp-2020,Pechenezhskiy_2020,deNeeve-2022,Eickbusch-2022,Siva-gkp-2023,LachanceQuirion-2023,Manset_2025}, opening the possibility for autonomous error correction at the physical qubit level; this could drastically reduce hardware requirements for quantum computers. This approach  would gain further impact with support of non-Clifford gates.

In this letter, we propose a protocol for an 
exponentially protected non-Clifford $\sqrt{T}$ gate on a self-correcting GKP qubit in a driven, dissipative superconducting circuit \cite{gkp-paper}. The protocol is based on controllably coupling ancillary Josephson Junctions to the qubit circuit to realize an effective $\varphi^4$ flux potential that implements a protected $\sqrt{T}$ gate through a phase revival mechanism analogous the the ones described in Ref.~\cite{gkp-paper,GKP}. Through analytics and numerics, we show that the resulting gate is topologically robust. 
In this way, our findings, together with the architecture proposed in Ref. \cite{gkp-paper}, realize an exponentially robust qubit with a universal set of protected single-qubit gates. 

\begin{figure}[t!]
    \centering
    \includegraphics[width=0.98\linewidth]{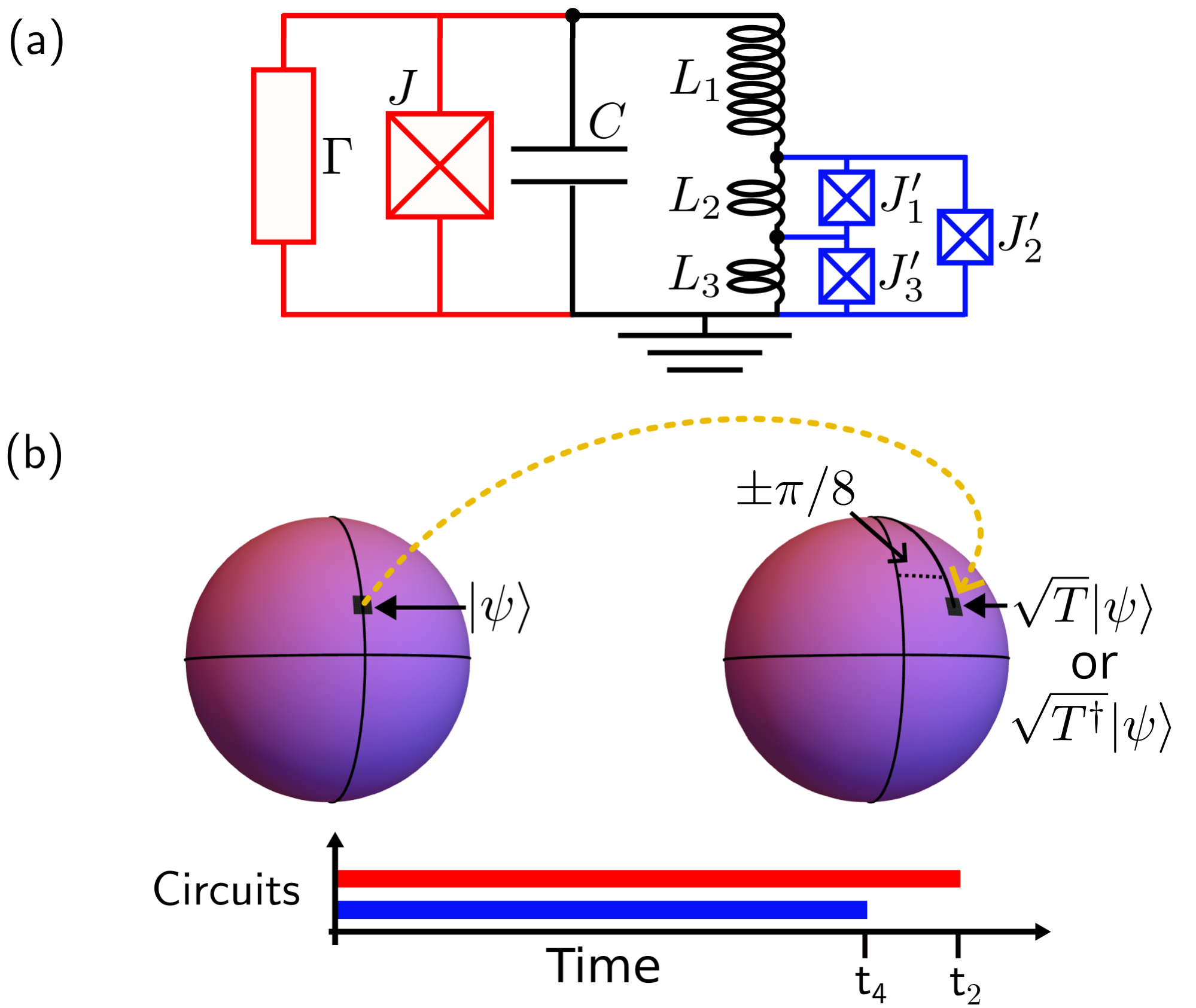}

    \caption{
    (a) Architecture for  generating an exponentially  protected $\sqrt{T}$ gate. The circuit consists of an LC resonator (black) connected via time-dependent switches to a Josephson Junction and dissipative element (red), and to ancillary Josephson junctions whose locations and junction energies are chosen to effectively generate a pure $\varphi^4$ potential (see main text). (b) Driving protocol for switches: blue and red subcircuits of panel (a) are activated for $t_4$ and $t_4+t_2$, respectively (see main text for definitions). 
    During the protocol, the system leaves and then reenters the codespace (orange line), with the logical state rotated by $\pm\pi/8$ (depending on the sign of the $\varphi^4$ potential) on the Bloch sphere (purple), which constitutes a $\sqrt{T}$ (+) or $\sqrt{T^{\dagger}}$ (-) gate.
    }
    \label{fig:circuit_protocol}
\end{figure}

\textit{Self-correcting GKP qubit---}
We first review the relevant features of the GKP qubit from Ref.~\cite{gkp-paper} that supports the  $\sqrt{T}$ gate. 
The device consists of an LC resonator with impedance $\sqrt{L/C} \approx h/2e^2 \approx  12.91\,{\rm k}\Omega$, connected via a  switch to a Josephson junction with Josephson energy $J$ and, capacitively, to a dissipative element---herein termed the resistor---with effective loss rate $\Gamma$; see Fig. \ref{fig:circuit_protocol}(a). Without the resistor, and with the switch closed, the circuit is  described by
\begin{equation}
    \HLCJ = \frac{\varphi^2}{2L} + \frac{q^2}{2C} - J \cos\left(\frac{2\pi \varphi}{\varphi_0}\right)\,,
    \label{eqn:H_LCJ}
\end{equation}
where $\varphi$ and $q = -2ie\partial_{\varphi}$ are the flux and charge quadratures for the resonator mode, respectively. The resistor is included by adding a term of the form $qQ_{\text{R}}/C_{\text{R}}$, where $Q_{\text{R}}$ is the Gaussian quantum noise signal from charge fluctuations on the resistor-side of the capacitative coupler $C_R$. 
This term adds an effective dissipator in the master equation for the system~\cite{gkp-paper,ULE_paper,breuer_theory_2007,gardiner_quantum_2004}. 
Deactivating the switch decouples the Josephson junction and resistor from the resonator, setting $J,(C_{\rm R})^{-1} \to 0$.


{
Under a suitable modulation of the switch (see below),  
the resonator mode spontaneously relaxes into a subspace of GKP states that encode a protected qubit~\cite{gkp-paper}. The subspace, termed the GKP codespace, is formed by the mutual high-eigenvalue subspace of the two commuting \textit{GKP stabilizers} $S_1 = \cos\left({2\pi \varphi}/{\varphi_0}\right)$ and $S_2 = \cos(2\pi  q/e)$, where the mode's $\varphi$ and $q$ support is confined near integer multiples of $\varphi_0$ and $e$, respectively.}
The qubit is encoded in the parity of these integers, with  logical operators 
\begin{equation} 
\sigma_z = {\Xi(\varphi/\varphi_0)},\quad 
\sigma_x = {\Xi(q/e)}, \quad
\sigma_y = -i \sigma_z \sigma_x\,.
\label{eqn:logicals}
\end{equation}
Here, the {\it crenellation function} $\Xi(x) = \sgn[\cos(\pi x)]$ returns the parity of the closest integer. 
Essentially, $\sigma_z$ captures whether the system is in an even or odd well of the flux potential of $H_{\rm LCJ}$, $\frac{\varphi^2}{2L} - J \cos\left(\frac{2\pi \varphi}{\varphi_0}\right)$, while $\sigma_x+i\sigma_y$ captures the relative  amplitude between different-parity wells.
Crucially, 
$\{\langle\sigma_i\rangle\} $ can only change if probability support leaks across either $\varphi = [n+1/2]\varphi_0$ or $q=[n+1/2]e$ for $n\in \mathbb Z$. 
Hence, $\{\sigma_i\}$ are invariant under any local perturbation of the system that keeps it confined in the codespace, making the qubit {\it protected}.

To  understand the relaxation into the GKP codespace, note that when the switch is connected---referred to as the \textit{stabilizer segment}---and for $J \gg h\fLC, k_{\rm B}T$,  with $T$ the resistor temperature and $\fLC = (2\pi\sqrt{LC})^{-1}$ the LC frequency, the resistor relaxes the system 
into the wells of the flux potential,
confining the system in the high-eigenvalue space of $S_1$. 
If the switch is then deactivated for a duration $1/4\fLC$---termed the \textit{free segment}---the system undergoes a $\pi/2$ rotation in phase space, which maps  $\varphi/\varphi_0$ to  $q/e$ and vice versa (up to a sign), thereby exchanging $S_1$ and $S_2$. 
The ensuing stabilizer segment relaxes the system again, leading to $\langle S_1\rangle \approx 1$.
Crucially, as we show below, setting the stabilizer segment duration to $\trev = \sqrt{LC}$ also ensures that 
$\langle S_2\rangle \approx 1$ at the end of the second stabilizer segment, ensuring full confinement in the code subspace.

To understand why $\langle S_2\rangle \approx 1 $ at the end of the stabilizer segment, first 
consider the  low-energy eigenstates  of $H_{\rm LCJ}$: these resemble harmonic oscillator eigenstates,  with characteristic decay length $\lambda_0\varphi_0$, where $\lambda_0 = [h\fLC/(4\pi^3 J)]^{1/4}$, and excitation energy $\varepsilon_0 = \sqrt{4e^2J/C}$~\cite{gkp-paper}, centered in the {wells} of the flux potential, $\varphi = N\varphi_0$ for $N\in \mathbb Z$. 
The eigenstate corresponding to the $\nu$th  excitation of well $N$, $\ket{N,\nu}$, 
 is effectively localized in well $N$ for   $\nu\lesssim (\pi/\lambda_0)^2$,
and has energy $E_{N,\nu} \approx  \nu \varepsilon_0 + N^2 \varepsilon_L$, 
where   $\varepsilon_L = \varphi_0^2/2L$. 
The GKP codespace is spanned by phase-coherent superpositions of these states, of the form $ \sum_{n,N}c_{n,N}|N,\nu\rangle$ with $c_{N,\nu}\approx c_{N+2,\nu}$ and $c_{N,\nu}\approx 0$~\cite{gkp-paper} for $\nu \gtrsim(\pi/\lambda_0)^2$, ensuring $\langle S_2\rangle \approx 1$ and $\langle S_1\rangle \approx 1$, respectively.
The parity of $N$ controls the $\sigma_z$ eigenvalue, since  $\sigma_z|N,\nu\rangle = (-1)^N|N,\nu\rangle +\mathcal O(e^{-(\pi/\lambda_0)^2})$.

Next, recall that $\langle S_2(0)\rangle \approx 1$ at the onset of the second stabilizer segment $(t=0)$. Expanding the evolution of the system as $|\psi(t)\rangle = \sum_{n,N}c_{n,N}(t)|N,\nu\rangle$~\footnote{Note that we are able to describe the state of our system with state vectors---even in the presence of dissipation---thanks to the SSE formalism.}, and noting $\avg{S_2(t)} \approx \sum_{N,\nu}\text{Re}[c_{N+2,\nu}^*(t)c_{N,\nu}(t)]$, we  conclude $c_{N,\nu}(0)\approx c_{N+2,\nu}(0)$. 
During the stabilizer segment, the coefficients acquire dynamical phases: $c_{N,\nu}(t) 
\approx c_{n,\nu}(0)^{i[\nu\varepsilon_0+N^2\varepsilon_L] t/\hbar}$. 
The $N^2$ part of the dynamical phase initially causes $\avg{S_2(t)} $ to decay to zero, taking the system out of the codespace~\footnote{This is the principal mode of decoherence when $\lambda_0 \ll 1$, since---crucially---the  resistor-induced dissipation  preserves inter-well coherence up to exponentially suppressed corrections~\cite{gkp-paper}}. 
Since  $N^2-(N+2)^2\in 4\mathbb Z$, the phases in  wells of the same parity {\it re-align} at multiples of the \textit{revival time} $\trev = \pi \hbar / 2\varepsilon_L$. 
At these instances, $c_{N,\nu}(t)\approx c_{N+2,\nu}(t)$, implying that the system returns to the high-eigenvalue subspace of $S_2$. 
The confinement persists when accounting for weak variations of eigenstates, spectra, and dissipation among the different wells; these modifications constitute phase-space local perturbations that do not alter the encoded logical state of the protocol if  $\lambda_0\ll 1$~\cite{gkp-paper}.
Thus, setting the stabilizer segment duration to  an integer multiple of $\trev$ ensures that 
 2 stabilizer segments confine the system in a state with $\langle S_1\rangle \approx \langle S_2\rangle \approx 1$~\cite{gkp-paper}. 
The alternating relaxation above 
 effectively resets  displacements induced by phase-space local noise, thus realizing \textit{dissipative error correction} (DEC) \cite{barnes_automatic_2000,Kitaev-memory-anyons,Terhal-QEC-quantum-memories-Review,brown_quantum_2016,deNeeve-2022,LachanceQuirion-2023,Li-Lieu-Dissipative-transitions,Lev-Arcady-Phillipe-GKP-Proposal,Max-Frederik-GKP}. 
The intrinsic stability afforded by 
DEC even protects the qubit against  imperfections in the protocol, such as mistargeting of circuit parameters and mistiming of the segment durations. 

Furthermore,  
the phase revival mechanism above generates a {\it protected $S^{\dagger}$ gate}: since $N^2 (\!\mod 4)=N (\!\mod 2)$, and $\varepsilon_L =  \pi \hbar / 2\trev$~\footnote{The latter equality follows using the condition $\sqrt{L/C} = h/2e^2$},
  we have $e^{-i\varepsilon_L N^2 \trev/\hbar}=(-i)^{N\mod 2}$~\cite{gkp-paper}. Thus, at $t=\trev$, wavefunction components in odd wells have gained a phase of $-i$ relative to components in even wells, $c_{N,\nu}(\trev)=(-i)^Nc_{N,\nu}(0)$; this is equivalent to the action of  an $S^{\dagger}$ gate on the logical qubit~\footnote{Recall that $\sigma_z|N,\nu\rangle=(-1)^{N}|N,\nu\rangle$, implying $(-i)^{N \mod 2}|N,\nu\rangle = e^{-i\frac{\pi}{2}(\sigma_z-1)}|N,\nu\rangle$.}.
Importantly, finite perturbations of the protocol   generate a continuous flow of the final state in phase space that do not induce logical errors when sufficiently small. 
In this sense, the gate is  \textit{topologically robust}. 





\textit{Protected $\sqrt{T}$ gates via $\varphi^4$ potential---}The phase revival mechanism  above can be leveraged to realize a topologically protected non-Clifford gate,by supplementing the quadratic flux potential with a higher-order potential~\cite{GKP}. 
Here, we focus on a protected $\sqrt{T}$ gate implemented via an additional potential \textit{quartic} in the flux. 
For simplicity, we first consider replacing the quadratic potential in $\HLCJ$ 
by a pure quartic potential, resulting in the Hamiltonian 
\begin{equation}
    H_4 = \frac{q^2}{2C} - J \cos\left(\frac{2\pi \varphi}{\varphi_0}\right)+ \frac{\epsilon_4}{\varphi_0^4} \varphi^4\,, 
    \label{eqn:H_phi4}
\end{equation}
with $\epsilon_4 < 0$.
For small enough $\epsilon_4$  and $\lambda_0$, the low-lying eigenstates of $H_4$  effectively coincide with $\{|N,\nu\rangle\}$, while their energies become  $E_{N,\nu} \approx \nu\varepsilon_0 + N^4 \epsilon_4$~\footnote{As with the $S^{\dagger}$ gate, corrections from the finite modifications do not take the system out of the code subspace when $\epsilon_4$ and $\lambda_0$ are small enough.}. 
The $N^4$ term leads to a \textit{quartic} dynamical phase, $c_{N,\nu}(t)\approx c_{N,\nu}(0)e^{-i(\nu\varepsilon_0+N^4 \epsilon_4)t/\hbar}$, which has a revival mechanism analogous to that of the quadratic potential~\cite{gkp-paper}: 
\begin{equation}
    N^4 \, (\!\!\mod 16) = N \, (\!\!\mod 2). 
\end{equation}
Thus, the dynamical phases for like-parity wells align at  the \textit{quartic revival time},  $t_4 = \pi \hbar/8|\epsilon_4|$. 
Here, the phases for even and odd wells differ by $\pi/8$, since $e^{-iN^4 \varepsilon_4 t_4/\hbar}=e^{i\frac{\pi}{8}(N\mod 2)}$, 
equivalent to the action of a $\sqrt{T}$ gate \footnote{Technically speaking, we realize a $\sqrt{T}$ gate for $\epsilon_4 < 0$; $\epsilon_4 > 0$  realizes a $\sqrt{T}^{\dagger}$ gate.}. 
As with the $S^{\dagger}$ gate, the $\sqrt{T}$ gate is topologically robust---i.e., exponentially insensitive to variations of the gate protocol, provided these variations are weak enough to keep the state confined in the codespace.

The protocol above can be generalized to 
potentials  with both a quadratic and a quartic part, i.e., $H'_4=\HLCJ +(\varepsilon_4/\varphi_0^4) \,\varphi^4$, which we argue below can be realized by connecting ancillary junctions to the GKP qubit.
In this case, the dynamical phases are $c_{N,\nu}(t)\approx c_{N,\nu}(0)e^{-i(\nu\varepsilon_0+\varepsilon_L N^2+\varepsilon_4 N^4) t/\hbar}$.
Even with incommensurate $\varepsilon_L $ and $\varepsilon_4 $, we can leverage the revival mechanism for each separately as follows: first,  let the system evolve via $H_4'$  for a duration $t_4$. Then,  disconnect the quartic potential 
and evolve via \textit{only} $\HLCJ$ for a time $t_2 = 4\trev - (t_4 \mod 4\trev)$. 
These two evolution segments---which we call the $\varphi^4$ \textit{segment} and $\varphi^2$ \textit{segment}, respectively---together ensure phase coherence between like-parity wells ($\avg{S_2} \approx 1$), while imparting a relative phase of $
\pi/8$ between even and odd wells. 
Cycles of  the stabilization protocol---here termed \textit{cleanup steps}---can subsequently be applied to reset any erroneous displacement incurred during the $\sqrt{T}$ gate protocol.

\begin{figure*}
    \centering
    \includegraphics[width=0.99\linewidth]{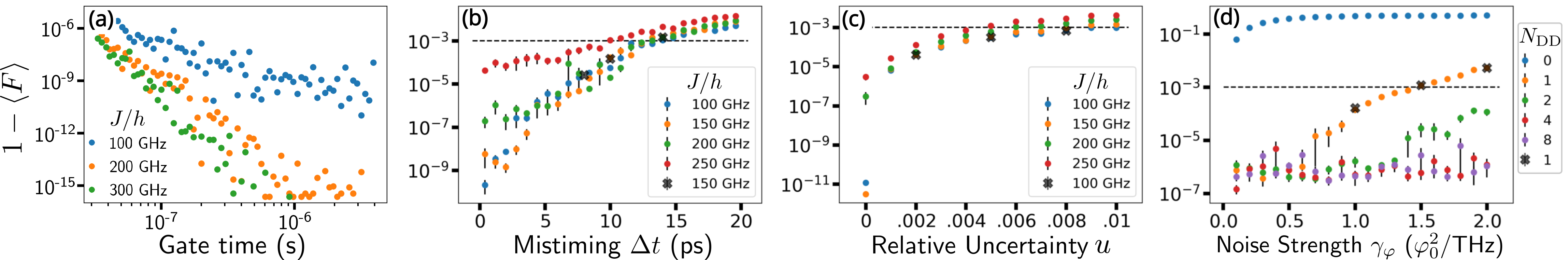}
    \caption{Simulated infidelities of $\sqrt{T}$ gate protocol---
    see main text for details. (a) Gate infidelity versus gate time with a pure $\varphi^4$ potential, 
    $\fLC=0.82\,{\rm GHz}$,
    no noise, and perfect control of the protocol and circuit parameters.  (b-d) Gate infidelity versus timing uncertainty $\Delta t$ (b), 
    relative uncertainty $u$ in circuit parameters (c), and flux noise strength $\gamma_{\varphi}$ (d), computed from an initial $\ket{X}$ logical state, using parameter set 3 of Table \ref{tab:circuit_params}, $T = 40$ mK, and $\Gamma = 1.5$ GHz. Data in panel (d) are obtained with  $J/h = 200$ GHz, and different numbers $\NDD$ of dynamic decoupling steps.  Dashed horizontal lines  indicate a gate infidelity of $10^{-3}$. We perform two additional cleanup steps, and average the results over (a,b) $10^4$ SSE trajectories, (c) $10^3$ SSE trajectories and 250 parameter mistargetings, and (d) 250 SSE trajectories and 300 noise signals. 
    Black crosses are averaged over (b) $10^6$, (c) 25000, and (d) $10^4$ trajectories to check convergence of statistical averaging.}
    \label{fig:all-sims}
\end{figure*}

The gate time of the protocol,  $\tGate \equiv t_4 + t_2$,  scales as $|\epsilon_4|^{-1}$,  and can be reduced until the 
$\varphi^4$ potential perturbs the eigenstates and excitation energies enough to differ significantly between wells  of $H_4'$. In the Supplementary Material (SM)~\cite{Supp_Info}, we estimate this minimal gate time to be of order $ \sim 10/\fLC$. 
Illustrating this, in   Fig. \ref{fig:all-sims}(a) we plot  gate infidelity (defined below)
versus gate time with $\fLC = 0.82$ GHz (corresponding to experimentally achievable $L=2.5\,\mu$H~\cite{Pechenezhskiy_2020,Manset_2025}) and for several values of $J$,  in the absence of any noise. 
Our simulation demonstrates lower gate infidelity with increased $\tGate$, consistent with our discussion above. Indeed, with no uncertainty in circuit parameters/control nor noise, the protocol can maintain infidelities  below $10^{-6}$  down to  $\tGate\lesssim 100$ ns, suggesting circuit realizations of the protocol could operate at or near this timescale. 

\textit{Circuit Realization---}
The $\varphi^4$ potential needed for the protected $\sqrt{T}$ gate can be effectively realized 
by inserting ancillary Josephson junctions along the linear inductor of the circuit, as  in Fig.~\ref{fig:circuit_protocol}(a).
Provided the  higher modes of the circuit~\cite{Beenakker-Critical-Current-fluctuations,Furusaki-Takayanagi-Tsukada-Josephson-Effect-vs-length,Cleuziou-Carbon-Nonatoube,DellaRocca-Current-Phase-Atomic-Contacts,Sochnikov-Non-Sinusoidal-Current-Phase-TI,English-Non-Sinusoidal-Current-Phase-graphene,van-Woerkem-ABS-semiconductor-junctions,Spanton-Current-Phase-InAs-nanowires,Kringhoj-hybrid-S-Sm-S-qubits,Leblanc-Charge-4e-Ge-JJs,Luca-JJ-rhombus} do not shift near resonance with the qubit mode  or become  thermally activated due to the ancillary junctions, they can be adiabatically eliminated through  a Born-Oppenheimer type approximation~\cite{Adiabatic-elimination,Supp_Info}. 
This results in a Hamiltonian of the form $\Heff = \HLCJ + \Veff(\varphi)$, where  $\Veff(\varphi)$  is  found  by minimizing the energy of the multi-mode Hamiltonian with respect to the {ancillary junction fluxes}, as a function of $\varphi$---see SM \cite{Supp_Info} for details. Importantly, $\Veff(\varphi)$ is a nonlinear function of $\varphi$ determined by the Josephson energies and locations of the ancillary junctions. 
Taylor expanding  $\Veff(\varphi)=\sum_{k=0}^{\infty} V_k \varphi^k$, the pure quartic potential required for a $\sqrt{T}$ gate  can effectively be realized if 
$V_k$ is negligible within the envelope $\sim \varphi_0/\pi\lambda_0$ of the GKP states \cite{gkp-paper} for all  $k \neq 0,2,4$~\footnote{This is achievable due to the nonlinear dependence of the ancillary junction phases on $\varphi$, which can lead to cancellation of the unwanted powers of $\varphi$ when Taylor expanding the ancillary junction potentials.}.
Ancillary junction parameters realizing these conditions can be found via computational search routines~\cite{Supp_Info}.

As an example, we optimize ancillary junction parameters for the circuit illustrated in Fig. \ref{fig:circuit_protocol}(a), consisting of three ancillary  junctions inserted at two nodes along the inductor, connecting the two nodes together and each node directly to ground.
The ancillary junctions energies, $\{J_i'\}$ and node locations comprise five independent circuit parameters. Using numerical search routines, we identify five parameter sets whose effective potential is nearly a pure $\varphi^4$ potential within the desired range of $\varphi$, shown in  Tab. \ref{tab:circuit_params}~\footnote{Note that, in our computational search, we allow the Josephson energies $J'_i$ to be either positive or negative. This can be achieved in practice by threading $\pi$ fluxes through the  loops of the blue subcircuit of Fig. \ref{fig:circuit_protocol}(a).}. We also list simulated gate infidelity (defined next section) and gate time for these parameter sets. Our identified parameters indicate a tradeoff between gate fidelity and gate time, 
though parameter sets 3-5 all have similar fidelity within the resolvable precision \footnote{Note that, while all of our data appear to follow the general trend of exponentially suppressed fidelities, we can technically only resolve infidelities of $\mathcal{O}(1/N_{\text{traj}})$, where $N_{\text{traj}}$ is the number of SSE trajectories sampled. We show data at or below these thresholds in any case, to illustrate the general exponential trend.}. 

\begin{table}[]
    \centering
    \begin{tabular}{||c||c|c|c|c|c||}\hline \hline
    Set $\#$ & 1 & 2 & 3 & 4 & 5 \\\hline\hline
    $L_1\,[\mu$H] & 2.35 & 2.36 & 2.40 & 2.41 & 2.41  \\ \hline  
    $L_2\,[\mu$H] & .0696 & .0688 & .0556 & .0471 & .0469 \\  \hline  
    $L_3\,[\mu$H]& .0758 & .0699 & .0436 & .0369 & .0393 \\  \hline  
    $J_1'/h \,[$GHz] & .0238 & .0614 & .843 & 1.18 & .870\\\hline  
    $J_2'/h \,[$GHz] & .583 & .629 & -.289 & -.139 & -.107\\\hline  
    $J_3'/h \,[$GHz] & .0118 & .00872 & .0420 & .0550 & .0250 \\\hline  \hline
    $t_{\rm gate}$ [$\mu$s] & 3.41 & 4.19 & 9.42 & 12.5 & 16.5 \\ \hline  
    $1-\avg{F}$ & $2\! \times\! 10^{-3}$ & $6\! \times\! 10^{-5}$& $3\! \times\! 10^{-8}$ & $3\! \times\!10^{-7}$ & $3 \! \times\! 10^{-8}$  \\ \hline  \hline
    \end{tabular}
    \caption{
    Parameter sets for the circuit architecture of Fig. \ref{fig:circuit_protocol}(a) that can realize $\sqrt{T}$ gates, along with calculated gate times and infidelities (see main text). 
    The parameters can be extended by re-scaling. Gate times and infidelities in the bottom rows are obtained from simulations  with $J = 150$ GHz and two cleanup steps, using $10^6$ SSE trajectories. 
    Negative junction energies 
    can be realized by threading fluxes appropriately through the circuit~\cite{Note7}. Note that sets 1 and 2 have $J_i' > 0$ for all $i$, and thus do not require any additional fluxes.}
    \label{tab:circuit_params}
\end{table}

\textit{Numerical Results---}
We now numerically demonstrate the resilience of our protocol to extrinsic noise, along with device and protocol imperfections. We use
the circuit  in Fig. \ref{fig:circuit_protocol}(a), with parameters given by Set 3 of Tab. \ref{tab:circuit_params}, and  model the dissipative dynamics during the $\varphi^4$, $\varphi^2$, and stabilizer segments with the universal Lindblad equation (ULE)~\cite{ULE_Kirsanskas,Frederik_Phd_thesis,ULE_Nathan_2019,ULE_Davidovic,ULE_paper,ULE_Nathan_2024}, using the stochastic Schr\"{o}dinger equation (SSE)~\cite{Dalibard_1992,Carmichael_1993}. 
We initialize the system in a stabilized GKP state encoding a pure logical $|X\rangle =\frac{1}{\sqrt{2}}(|0\rangle +|1\rangle)$ state, with Bloch vector $\boldsymbol{\sigma}_0$. 
Using the expectation $\langle \boldsymbol \sigma\rangle_{\rm f}$ of the logical operators  in the final state resulting from the simulation, we compute the fidelity of the protocol as $F= (1+\langle \boldsymbol \sigma\rangle_{\rm f}\cdot \boldsymbol \sigma _{\rm target})/2$, with  $ \boldsymbol \sigma _{\rm target}$ representing the target logical state obtained by rotating $\boldsymbol{\sigma}_0$ by $\pm\pi/8$ around the $z$ axis
\footnote{Note that this definition is equivalent to the squared overlap $|\langle \Psi_{\text{target}}\lvert \Psi_{\text{f}}\rangle|^2$, where $\ket{\Psi_{\text{target}}}$ and $\ket{\Psi_{\text{f}}}$ are the logical wavefunctions for the target and final states with Bloch vectors $\boldsymbol{\sigma}_{\text{target}}$ and $\boldsymbol{\sigma}_{\text{f}}$, respectively.}.

We first probe the protocol's resilience against mistiming of the various segments: we  simulate evolution under the protocol where each $\varphi^4$ and $\varphi^2$ segment has its length independently modified by some time $\delta t$ randomly chosen from $[-\Delta t/2,\Delta t/2]$. For any subsequent cleanup steps, we modify the stabilizer and free segment durations to ensure the total duration of each cleanup step is constant \footnote{This way, the frequency of the driving (which can be very precisely realized in real devices) is preserved, even if the details of each driving period are not.}. In Fig. \ref{fig:all-sims}(b) we plot gate infidelity $1-F$    
versus $\Delta t$. We see the protocol maintains gate fidelities of four or more nines up to timing uncertainty of $10$ ps or more for the simulated circuit. 
Recalling $L\sim 2.5 \,\mu{\rm H}$ and  that  timing tolerance is expected to scale with $ 1/\fLC\propto L$; 
circuit realizations with larger $L$~\cite{Manset_2025}, will thus increase the timing tolerance in proportionally to $L$, with $\Delta t_{\rm min} \sim 4 \, {\rm ps}\times L[\mu{\rm H}]$. 

We next probe the protocol's resilience  to imperfect targeting of circuit parameters, by drawing $\{L_i\}_i$ and $\{J'_i\}_i$ from normal distributions whose means are the values listed in Tab. \ref{tab:circuit_params}, and whose standard deviations are $u$ times that value, for various $u \geq 0$. Fig. \ref{fig:all-sims}(c) shows infidelity $1-F$ versus the relative uncertainty $u$. 
The gate protocol maintains $<10^{-3}$ infidelity for  $u\lesssim 1\%$, and is more stable for smaller $J$ (larger $\lambda_0$). 

We finally probe the protocol's resilience to flux noise~\footnote{The principal effects of charge noise (stochastic displacement along $\varphi$ in phase space) are immediately corrected by the resistor \cite{gkp-paper}. Flux noise, conversely, is not corrected until after a free segment.}.
We include flux noise in our simulations via a fluctuating offset $\xiphi(t)$ in the inductor flux:
$
    H \mapsto H + \frac{\xiphi(t)}{L}\varphi\,,
$
where $\xiphi(t)$ is a Gaussian signal with power spectral density (PSD) $S_{\varphi}(\omega) = \gamma_{\varphi}\,e^{-|\omega|/\Lambda}(2\pi\,\text{Hz})/(|\omega|+\omega_0)$.
This is the PSD for $1/f$ noise, with IR and UV cutoffs $\omega_0$ and $\Lambda$, respectively, while $\gammaphi$ parameterizes the noise strength. 
In our simulations, we take $\omega_0 = 10^{-4}\,$Hz and $\Lambda = 1\,$MHz, and vary the noise strength $\gammaphi$. See SM~\cite{Supp_Info} for  details on the generation of $\xi_{\varphi}(t)$.

To counter the effects of flux noise, we implement dynamic decoupling by breaking the $\varphi^4$ segment into $\NDD$ steps, with free segments of length $1/(2\fLC)$ between. This half-cycle evolution sends $\varphi \mapsto -\varphi$, effectively reversing the direction of noise-induced diffusion and canceling low-frequency noise. We choose the duration of the $\NDD$ steps according to the Uhrig dynamical decoupling (UDD) scheme \cite{Uhrig-DD}. We show the gate infidelity versus noise strength $\gammaphi$, for several different number of UDD steps $\NDD$, in Fig. \ref{fig:all-sims}(d). 
Evidently, even a few UDD steps are sufficient to suppress the effects of $1/f$ flux noise with strengths on the order of $\sim 1 \varphi_0^2/{\rm THz}$. 

Flux noise can also be present in the loops comprising the ancillary junctions. Noise in these loops will modify the Josephson potential of the ancillary junctions, leading to corrections $\delta V_k(t)$ to the coefficients $V_k$ of the effective potential. For $1/f$ flux noise, such corrections will be slowly-varying; the corrections $\delta V_k(t)$ with $k$ odd will be suppressed by dynamical decoupling, and $k$ even can be treated as a form of parameter mistargeting, akin to that described above. {Based on this intuition, we estimate a tolerance for fluxes of up to $\sim .01 \varphi_0$ through each of the small loops \footnote{This estimate can be obtained by treating flux noise as a quasistatic correction $\varphi \mapsto \varphi + \delta\varphi$, and using the angle addition formula
\begin{equation*}
\begin{split}
\cos\left(\frac{2\pi}{\varphi_0}(\varphi+\delta\varphi)\right) =& \cos\left(\frac{2\pi}{\varphi_0}\delta\varphi\right)\cos\left(\frac{2\pi}{\varphi_0}\varphi\right)\\
&\quad - \sin\left(\frac{2\pi}{\varphi_0}\delta\varphi\right)\sin\left(\frac{2\pi}{\varphi_0}\varphi\right)
\end{split}\,.
\end{equation*}
We can then relate $\delta \varphi$ to the effective mistargeting $u$ via $\delta\varphi = \varphi_0\arccos(1-u)/2\pi$. For $u = .002$ [c.f. Fig. \ref{fig:all-sims}(c)], this yields $\delta\varphi \approx .01\varphi_0$.}.}
Thus, we do not expect flux noise in the small loops to pose any serious additional issues. 


\textit{Discussion---}
Here we introduced a protocol for high-fidelity $\sqrt{T}$ gates on GKP qubits in a driven, dissipative superconducting circuit. We proposed a family of circuits that can realize our gate protocol, and have verified through simulation the protocol's resilience to a variety of errors. In line with our prior results, the resistor corrects errors from phase-space local noise, leading to exponentially suppressed gate infidelities. 
GKP proposed a similar mechanism for $T$ gates with a $\varphi^3$ potential \cite{GKP}. In our case, using a $\varphi^4$ potential  allows us to realize $\sqrt{T}$ gates, as well as perform dynamical decoupling. The confinement from the capacitor and junction potential also allow us to avoid issues with the rapid variation of higher-order potentials in finitely squeezed GKP states \cite{GKP-cubic-phase-gate-unsuitability}. 

Like the stabilization protocol, the $\sqrt{T}$ gate protocol described here requires a switch mechanism that can suppress Josephson couplings by orders of magnitude on the order of $10-50$  picoseconds, for current or near-term resonators~\cite{Manset_2025}. Realizing such a switch is challenging, but carries significant reward: a non-Clifford $\sqrt{T}$ gate, together with the Clifford gates and readout scheme in Ref. \cite{gkp-paper}, mean this device is a potential platform for \textit{universal}, robust single-qubit computation. Even if not within the threshold for self-correcting logic, our gate protocol still has promise for generating magic states.  

Due to the large separation of timescales, $t_4 \gg \trev$, it is  possible that the ancillary circuit can be left connected at all times. The effective potential $\Veff$ from the ancillary junctions, being a finite-order polynomial in $\varphi$, can be treated as a phase-space local noise source during the native stabilization protocol; errors incurred by leaving the ancillary circuit connected are thus correctable \cite{GKP}. This insight has the potential to significantly simplify the control required for our device.


An interesting future direction is to explore whether multi-qubit gates can be implemented using a similar revival mechanism. Together with the $\sqrt{T}$ gate protocol here and our prior results, this architecture could realize universal self-correcting logic on any number of qubits. In addition, 
topologically-protected gate protocols have also been explored for other bosonic qubits, such as the $0-\pi$ qubit~\cite{kitaev2006protectedqubitbasedsuperconducting,bkp,Kolesnikow-Smith-Thomsen-Alase-Doherty-0-pi}; exploring analogues of our protocol for this platform too will be another interesting future direction.


\textit{Acknowledgements---}
We are grateful to Alexei Kitaev, Matthew Matheny, and Anders S\o{}rensen for helpful discussions. F.N. was supported by the U.S. Department of Energy, Office of Science, Basic Energy Sciences under award DE-SC0019166, the Simons Foundation under award 623768,  the Carlsberg Foundation, grant CF22-0727, and the Novo Nordisk Foundation, Grant No. NNF22SA0081175, NNF Quantum Computing Programme. G.R. is grateful for support from the Simons Foundation, as well as support from the NSF DMR Grant number 1839271, and from the IQIM, an NSF Physics Frontiers Center. 
This work was performed in part at the Aspen Center for Physics, which is supported by National Science Foundation grant PHY-1607611. 

\bibliography{bibliography_T_gate.bib}

\end{document}


\title{Supplementary Information for ``Exponentially robust non-Clifford gate in a driven-dissipative circuit"}
\author{Liam O'Brien, Gil Refael, Frederik Nathan}
\maketitle

\setcounter{equation}{0}
\setcounter{figure}{0}
\setcounter{table}{0}
\setcounter{page}{1}
\makeatletter
\renewcommand{\theequation}{S\arabic{equation}}
\renewcommand{\thefigure}{S\arabic{figure}}
\renewcommand{\bibnumfmt}[1]{[S#1]}
\renewcommand{\citenumfont}[1]{S#1}


\section{Obtaining Effective Potentials $V_{\rm eff}$ from Circuits}
In this appendix, we outline how to obtain the effective potential $V_{\rm eff}(\varphi)$ for a given circuit using a Born-Oppenheimer---or adiabatic elimination---approximation,  and identify the conditions under which this approximation is justified. Such a procedure is how we realize approximately quartic potentials with the circuit architecture of Fig. 1(a) of the main text.

The Hamiltonian for a circuit with arbitrarily configured ancillary Josephson junctions that have $N$ nodes in total,  can be written in terms of the node charges $\{q_n\}$ and fluxes $\{\varphi_n\}$ as 
\newcommand{\vphi}{\{\varphi_n\}}
\newcommand{\vq}{{\{q_n\}}} 
\begin{equation}
H= \frac{1}{2}\sum_{m,n=1}^N ({\boldsymbol C}^{-1})_{mn}q_m q_n +V(\varphi_1,\ldots, \varphi_N)\,,
\end{equation}
with $\boldsymbol{C}$ the mutual capacitance  matrix of the nodes, and $V(\vphi)$ the effective flux potential of the circuit as a function of all node fluxes, accounting for all inductances and Josephson energies.
Throughout our analysis, we let $q \equiv q_1$ and $\varphi\equiv \varphi_1$ denote the flux and charge at the node of the non-ancillary Josephson junction, henceforth termed the {\it main node}. 
Without loss of generality, we may write 
\begin{equation}
    V(\varphi,\varphi_2,\ldots \varphi_N)=-J\cos(2\pi \varphi/\varphi_0)+\frac{\varphi^2}{2L}+\Delta V(\varphi;\varphi_2,\ldots \varphi_N)\,,
\end{equation} 
with $L$ the inductance of the main node in the absence of all ancillary junctions. 
We also neglect the mutual capacitance between the main nodes and the ancillary nodes; any residual coupling can be incorporated as a quantum noise term, against which the qubit is resilient at sufficiently small strengths. Thus, we may write $\sum_{m,n=1}^N \frac{1}{2} ({\boldsymbol C}^{-1})_{mn}q_m q_n =\frac{q^2}{2C} + \sum_{m,n=2}^N \frac{1}{2} ({\boldsymbol C}^{-1})_{mn}q_m q_n$, such that 
\begin{equation}
H= \frac{q^2}{2C}+\frac{\varphi^2}{2L}-E_J\cos(2\pi \varphi/\varphi_0)+ \frac{1}{2}\sum_{m,n=2}^N ({\boldsymbol C}^{-1})_{mn}q_m q_n +\Delta V(\varphi;\varphi_2,\ldots \varphi_N)\,.
\label{eqa:hexp}
\end{equation}
Our goal is to integrate out the variables at all other nodes, i.e., $\varphi_2,\ldots \varphi_N$ and $q_2\ldots q_N$. 

As a first step, we introduce the functions $\bar \varphi_2(\varphi),\ldots \bar \varphi_N(\varphi)$, defined through 
\begin{equation}
    (\bar \varphi_2(\varphi),\ldots \bar \varphi_N(\varphi))=\arg\underset{\varphi_2,\ldots \varphi_N}{\min}\Big[\Delta V(\varphi;\varphi_2,\ldots \varphi_N)\Big]\,. 
\end{equation}
Note that $\bar \varphi_n(\varphi)$ is a  function of $\varphi$ only, and thus, as a quantum operator, acts only non-trivially on the Fock space associated with the main node. 
Next, we write 
\begin{equation}
    \Delta \varphi_n = \varphi_n-\bar \varphi_n(\varphi)\,.
\end{equation}
Note that $[q_m,\Delta \varphi_n(\varphi)]=-i\hbar\delta_{mn} $ for $m,n\geq 2$; hence $\{(\Delta \varphi_n,q_n)\}$ for $n\geq 2$ forms  a set of independent canonically conjugate quantum variable pairs. 

We now Taylor expand $V(\varphi;\varphi_2,\ldots \varphi_N)$ in powers of the quantum variables $\{\Delta \varphi_n\}$. 
Noting that $\varphi_n = \bar \varphi_n(\varphi)$ minimizes $\Delta V$,

\begin{equation}
\Delta V(\varphi;\varphi_2,\ldots \varphi_N) = V_0(\varphi)+\sum_{m,n=2}^NV_{mn}^{(2)} (\varphi)\Delta \varphi_m\Delta \varphi_n+\mathcal O(\Delta \varphi^3)\,. 
\end{equation}
With $V_0(\varphi)=\underset{\varphi_2,\ldots \varphi_N}\min \Delta V(\varphi;\varphi_2,\ldots \varphi_N)$, and $V^{(2)}_{mn}(\varphi)$ denoting the second order Taylor coefficients.
Inserting this in Eq.~\eqref{eqa:hexp}, we find 
\begin{equation}
H = \frac{q^2}{2C}+\frac{\varphi^2}{2L}-E_J\cos(2\pi\varphi/\varphi_0)+ V_0(\varphi)+\sum_{m,n=2}^{N} \left(\frac{1}{2}(\boldsymbol{C}^{-1})_{mn} q_m q_n +V^{(2)}_{mn}(\varphi)\Delta \varphi_m\Delta \varphi_n)\right)+\mathcal O(\Delta \varphi^3)\,.
\label{eqn:Hamiltonian_expanded_second_order}
\end{equation}
The penultimate term in the sum above is a system of coupled harmonic oscillators that can be solved in terms of its normal modes, and has frequencies $\{\omega_k(\varphi)\}$; we refer to these modes as \textit{higher modes}. Below we perform an example calculation for their frequencies for the circuit depicted in Fig. 1(a) of the main text. 
When the frequencies of the higher modes, $\{\omega_k\}$ are much larger than the energy scales of the first term, drive protocol, and  temperature, we can assume the system to effectively be in the ground states with respect to these terms, through an adiabatic elimination procedure~\cite{Adiabatic-elimination}. This approximation is justified even when accounting for the higher-order nonlinear corrections above, as long as they are not strong enough to significantly weaken the excitation gap.
In this case, we can replace the system of coupled oscillators in \eqref{eqn:Hamiltonian_expanded_second_order} above with the vacuum energy of the higher modes, $\varepsilon_0(\varphi)=\frac{1}{2}\hbar \sum_k \omega_k(\varphi)$, leading to 
\begin{equation}
    H_{\rm eff} \approx  
 \frac{q^2}{2C}+\frac{\varphi^2}{2L}-E_J\cos(2\pi\varphi/\varphi_0)+ V_{\rm eff}(\varphi)\,, 
\end{equation}
where
\begin{equation}
V_{\rm eff}(\varphi)=\underset{\varphi_2,\ldots \varphi_N}\min V(\varphi,\varphi_2,\ldots \varphi_N)+\frac{\hbar}{2}\sum_k\omega_k(\phi)\,.
\label{eqn:V_eff_minimization}
\end{equation}
This is what we wanted to show. 
We ignore the quantum fluctuation terms in our analysis below, since we assume it to be negligible.

\subsection{Numerical search routine}
In our simulations, we obtain the effective potential as a function of $\varphi$ by numerical minimization of the right hand side of \eqref{eqn:V_eff_minimization}. 
 
We can then extract the coefficients $V_k$ by fitting the extracted $V_{\text{eff}}(\varphi)$ to a polynomial in $\varphi$.
Having access to the coefficients allows us to perform a numerical search of the parameter space, as described in the main text. More precisely, we can search for local maxima of the fourth-order coefficient $V_4$, whilst checking that the higher-order coefficients $V_6,V_8,\ldots$ are small enough---namely, that they obey the bounds \eqref{eqn:coefficient_constraint} and \eqref{eqn:coeff_condition_FT} identified in subsequent sections below \footnote{Note that the effective potential given by \eqref{eqn:V_eff_minimization} will be even in $\varphi$, and so all $V_k$ with $k$ odd vanish.}. This method was used to extract the five parameter sets identified in Tab. I of the main text.

\subsection{Example: calculation of higher-mode frequencies for circuit in Fig. 1(a)}
\label{sec:finite_JJ_cap}

\begin{figure}
    \centering
    \includegraphics[scale=0.3]{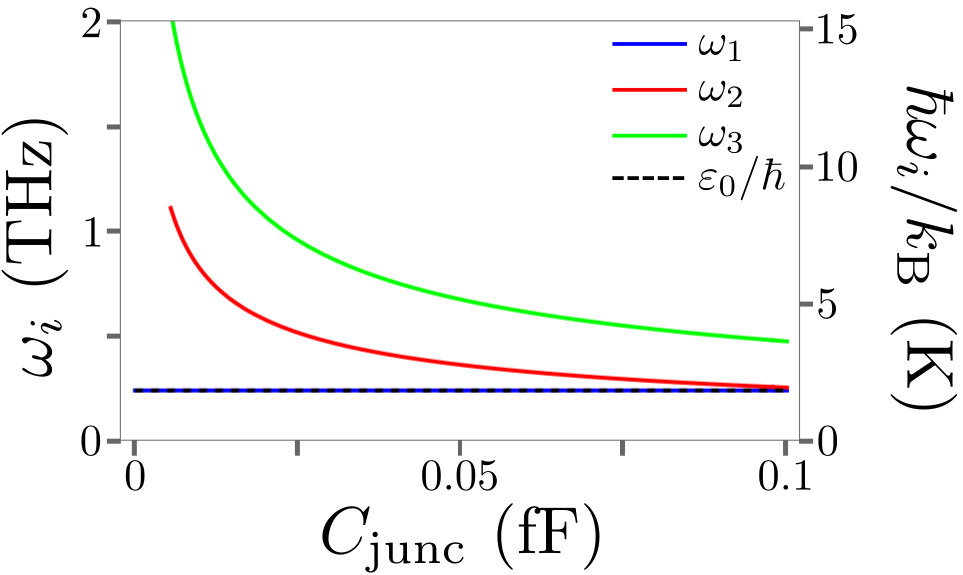}
    \caption{Normal mode (angular) frequencies $\omega_i$ as a function of ancillary junction capacitance $C_{\text{junc}}$. Left and right vertical axes show the frequencies in THz and equivalent temperature in Kelvin, respectively. Black dashed line (where visible) indicates the transition frequency $\varepsilon_0/\hbar = \frac{1}{2\pi}\sqrt{2J/(h\sqrt{LC})}$ of the low-lying eigenstates of $\HLCJ$. Frequencies were calculated by solving the linearized system in equations (\ref{eqn:E_L_eqns}a-c), with $C_i \equiv C_{\text{junc}}$ for all junctions, circuit parameters given by parameter set 3 of Tab. I in the main text, and $J/h = 150$ GHz.}
    \label{fig:frequencies}
\end{figure}

Here, we show an example of the procedure above, and calculate the higher mode frequencies for the circuit in Fig.~1(a) of the main text. In doing so, we identify under which conditions the adiabatic elimination procedure described above is justified. 

Our calculation accounts for the finite capacitances of the ancillary junctions. 
The capacitance for the main junction (i.e., the junction at the main node) is in parallel with the capacitance $C$ of the LC resonator, and so does not create any new resonator modes.
Letting $\varphi_1',\varphi_2'$ denote the phases at the nodes between $L_1$ and $L_2$, and $L_2$ and $L_3$ (respectively), the Lagrangian of the circuit (without the resistor) is
\begin{equation}
    \begin{split}
        \mathbb L &= \frac{C \dot{\varphi}^2}{2} + J\cos\left(\frac{2\pi \varphi}{\varphi_0}\right) - \frac{(\varphi - \varphi_1')^2}{2L_1} - \frac{(\varphi_1'-\varphi_2')^2}{2L_2} - \frac{\varphi_2'^2}{2L_3} + J_1'\cos\left(\frac{2\pi(\varphi_1'-\varphi_2')}{\varphi_0}\right) \\
        &\qquad + J_2'\cos\left(\frac{2\pi \varphi_1'}{\varphi_0}\right) + J_3'\cos\left(\frac{2\pi \varphi_2'}{\varphi_0}\right) + \frac{C_1'}{2}(\dot{\varphi_1'}-\dot{\varphi_2'})^2 + \frac{C_2'}{2}\dot{\varphi_1'}^2 + \frac{C_3'}{2}\dot{\varphi_3'}^2
    \end{split}\,.
    \label{eqn:JJ_cap_lagr}
\end{equation}
Assuming $L_1 \gg L_2,L_3$, the phases $\varphi_1',\varphi_2'$ will be close to ground, and we can expand the cosine potentials from the ancillary Josephson junctions according to $J_i'\cos(2\pi \varphi_i'/\varphi_0) \approx 1 - 2\pi^2 J_i' \varphi_i^2/\varphi_0^2$. For states confined to the codespace, we will also have $\varphi \approx N\varphi_0$ for integer $N$; we can thus expand the main Josephson potential according to $J\cos(2\pi \varphi/\varphi_0) \approx 1 - 2\pi^2 J(\varphi-N\varphi_0)
^2/\varphi_0$. With these approximations, the Euler-Lagrange equations read:
\begin{subequations}
    \begin{align}
        C\ddot{\varphi} &= -\frac{4\pi^2 J}{\varphi_0^2}(\varphi-N\varphi_0) - \frac{\varphi-\varphi_1'}{L_1}\,, \\
        C_1'(\ddot{\varphi_1'}-\ddot{\varphi_2'}) + C_2'\ddot{\varphi_1'} &= -\frac{4\pi^2}{\varphi_0^2}\left[J_1'(\varphi_1'-\varphi_2') + J_2'\varphi_1'\right] + \frac{\varphi-\varphi_1'}{L_1} - \frac{\varphi_1-\varphi_2}{L_2}\,, \\
        -C_1'(\ddot{\varphi_1'}-\ddot{\varphi_2'}) + C_3'\ddot{\varphi_2'} &= -\frac{4\pi^2}{\varphi_0^2}\left[-J_1'(\varphi_1'-\varphi_2')+J_3\varphi_2'\right]+ \frac{\varphi_1-\varphi_2}{L_2} - \frac{\varphi_2}{L_3}\,.
    \end{align}
    \label{eqn:E_L_eqns}
\end{subequations}
These equations define a linear system $ \mathbf{C}\ddot{\vec{\varphi}} = \mathbf{V}\vec{\varphi}  + \vec{b}$ of differential equations, where $\vec{\varphi} = \begin{pmatrix} \varphi & \varphi_1' & \varphi_2'\end{pmatrix}^T$ and $\vec{b} = \begin{pmatrix} 4\pi^2 J N/\varphi_0 & 0 & 0 \end{pmatrix}^T$. 
Note that the presence of $\vec{b}$ only adds a constant (in time) shift to the solution vector $\vec{\varphi}$, so we set $\vec{b} = 0$ without loss of generality. To solve the remaining linear system, we look for the normal modes of the system; we obtain three frequencies $\omega_1$, $\omega_2$, $\omega_3$ corresponding to $\sqrt{-e_1}$, $\sqrt{-e_2}$, $\sqrt{-e_3}$, where the $e_i$ are the eigenvalues of the matrix $\mathbf{C}^{-1}\mathbf{V}$. These normal mode frequencies are expected to correspond to the transition frequencies of the oscillator modes when the system is quantized (i.e., $\omega_k(\varphi)$ in Eq. \eqref{eqn:V_eff_minimization} above); we want to check that the two larger frequencies remain frozen out at the operating temperatures of the device.

We compute the three normal mode frequencies for the parameters in set 3 of Tab. I of the main text, with $J/h = 150$ GHz. For simplicity, we take the capacitance for all three ancillary junctions to be the same: $C_i' \equiv C_{\text{junc}}$. The three frequencies are shown in Fig. \ref{fig:frequencies}. Evidently, there is a substantial gap between $\omega_1$ and $\omega_2$, $\omega_3$ for junction capacitances $C_{\text{junc}} \lesssim 0.1 $fF. Here mode $1$ encodes the qubit, with  frequency (when the switch is connected) given by $\omega_1 \approx \frac{\hbar}{\varepsilon_0}= \frac{1}{\sqrt{L_JC}}$ with $L_J = \varphi_0^2 /4\pi^2 E_J$ the effective Josephson inductance of the main node.
The gap can be increased further by lowering the main node junction energy, which will reduce $\omega_1$ while not significantly affecting $\omega_2$ or $ \omega_3$. 
Furthermore, all frequencies have equivalent temperatures on the order of ones to tens of Kelvin, meaning the higher frequencies would be frozen out at operating temperatures on the order of hundreds of milliKelvin or lower (where our device is expected to perform well \cite{gkp-paper}), assuming the control of the circuit is adiabatic on this scale. Adiabatically eliminating the excited states of these other resonator modes  is thus justified in these cases.

The $0.1\,{\rm fF}$ threshold on $C_{\text{junc}}$ may be achievable with current technology. For junctions with specific capacitances of $10-100\,\text{fF}/\mu\text{m}^2$ (which has been demonstrated experimentally \cite{Deppe-Saito-Tanak-Takayangi-JJ-cap,Yeh-et-al-JJ-cap}), this condition would require a junction area of $\lesssim .01-.001\,\mu$m$^2$.
Junctions this small are challenging to fabricate, but have been realized \cite{Improved-JJ-reproducibility,Gate-tunable-JJ}.
Likewise, Ref.~\cite{anferov2025millimeterwavesuperconductingqubit} reported fabrication of Junctions with capacitance-to-Josephson energies in the range of $\sim 0.1 {\rm fF}/h {\rm GHz}$, which is compatible with the threshold above, and the required junction energies of  $.01-.8\, h{\rm GHz}$ listed in Table 1 in the main text.
Of course, it is not immediately clear that the scaling of capacitance per unit area would hold for the small junctions we require; a more definitive answer would require detailed electromagnetic calculations involving the exact device geometry, which is beyond the scope of this current work. Nonetheless, even if the junction capacitances lie at or near this threshold, it may still be the case that these additional modes are not thermally activated. For the example of Fig. \ref{fig:frequencies}, the excitation energies of the additional modes have equivalent temperatures on the order of ones of Kelvin, well above the expected operating temperature of the device. In such a case, the excited states of \textit{all} resonator modes---including the one encoding the qubit---are frozen out.

\section{Conditions on the effective potential}
\label{sec:ham_analysis}

In this appendix, we infer the conditions on the effective potential in order for it to generate a high-fidelity $\sqrt{T}$ gate. 
To that end, we consider a Hamiltonian of the form
\begin{equation}
    H = \HLCJ + \sum_{k=0}^{\infty} V_k \left(\frac{\varphi}{\varphi_0}\right)^k\,,
\label{eqn:H_general}
\end{equation}
where 
\begin{equation}
    \HLCJ = \frac{\varphi^2}{2L} + \frac{q^2}{2C} - J\cos\left(\frac{2\pi\varphi}{\varphi_0}\right)\,.
\end{equation}
This is the Hamiltonian of the circuit drawn in Fig. 1(a) of the main text, with the $V_k$ determined by the details of the ancillary junctions [blue subcircuit in Fig. 1(a)]. Intuitively, the $k=4$ term is the one we want to realize the $\sqrt{T}$ gate protocol described in the main text; all other terms (except possibly $k=0,2$ \footnote{\label{footnote:V_2} The $k=0$ term is a constant shift of $H$, which has no meaningful dynamics. The $k=2$ term can be accounted for by altering the definition of $t_2$ in the main text to address the fact that the revival time for the quadratic phases is different during the $\varphi^4$ and $\varphi^2$ segments, i.e.,
$$t_2 = \left(4 - \hspace{-8pt}\mod_{\!\trev'} t_4\right)\trev\,,$$
where $\trev' = 1/(2\pi\fLC + 4V_2/h)$.}) act as unwanted corrections we wish to suppress.


Heuristically, a high-fidelity $T$-gate requires two key conditions.
\begin{enumerate}
    \item The eigenstates of $H_{\rm LCJ}$ are effectively unchanged relative to the $V_k \equiv 0$ case, to maintain phase coherence between wells.
    \item  The phases from the $k \neq 4$ terms neither take the system out of the codespace nor affect the logical information.
\end{enumerate}
We address each of these conditions in subsections~\ref{seca:tolerance} and \ref{seca:phases}, respectively.
The conditions above lead us to upper limits on the gate speed and on the magnitudes of the residual non-quartic coefficients of the effective flux potential. Specifically, we find that the non-quartic coefficients must satisfy
\begin{eqnarray}
V_k \ll \frac{\hbar}{\kappa^k t_4}\,,
\end{eqnarray}
where $\kappa \equiv \sqrt{\coth(2\varepsilon_0/\kb T)}/\pi\lambda_0$ [see Eq.~\eqref{eqa:cond2}], and that 
 the gate time $t_4$ must satisfy
\begin{equation}
t_4 \gg C/\fLC,
\end{equation}
with  $C$ a numerical constant of order $\gtrsim \mathcal O(1)$ [Eq.~\eqref{eqa:t4min}]. 


\subsection{Threshold for modification of eigenstates}
\label{seca:tolerance}
We first obtain the conditions ensuring that  the eigenstates $\{\ket{N,\nu}\}$ in each well are  effectively unchanged. This will lead to bounds on the allowed magnitudes of the coefficients $V_k$. The bound for $k=4$ will, in addition, bound the fastest achievable gate time.  

To bound the coefficients $\{V_k\}$ we first assume that the $\varphi_k$ operators do not hybridize the low-lying eigenstates of $H_{\rm LCJ}$  in different wells. For the qubits we consider, this will never be an issue, as long as the effective potential remains bounded \cite{gkp-paper}. In general, then, the perturbed eigenstates $\{\ket{N,\nu}'\}$ in each well are given in terms of the corresponding unperturbed states $\{|N,\nu\rangle\}$ via a unitary $U^N$
\begin{equation}
    \ket{N,\nu}' = \sum_{\nu}U^N_{\nu\mu}\ket{N,\mu}\,.
\label{eqn:unitary_gen}
\end{equation}
To zeroth order in $V_{\rm eff}$, $U^{N}=\mathbb{I}^{N}$, i.e., the identity in well $N$. The leading-order correction can be found (for instance, via a Schrieffer-Wolff transformation \cite{Schrieffer-Wolff}) to be
\begin{equation}
U^N_{\mu\nu} = \frac{1}{2}\sum_{k,k'} \frac{V_k V_{k'}}{\varphi_0^{k+k'}} \sum_{\eta}\frac{[\langle N,\mu\rvert \varphi^k \rvert N,\eta\rangle\langle N,\eta \rvert \varphi^{k'} \lvert N,\nu\rangle][2E_{N,\eta}-E_{N,\mu}-E_{N,\nu}]}{(E_{N,\eta}-E_{N,\mu})(E_{N,\eta}-E_{N,\nu})(E_{N,\nu}-E_{N,\eta})}\,.
\label{eqn:unitary_leading_order}
\end{equation}
These corrections will be small if:
\begin{equation}
\frac{V_k}{\varphi_0^k}\left|\frac{\bra{N,\mu}\varphi^k\ket{N,\eta}}{E_{N,\mu} - E_{N,\eta}}\right| \ll 1
\label{eqn:pert_condition}
\end{equation}
for all relevant $k$, $N$, $\mu$, and $\eta$. Let us now discuss when this condition is satisfied.

We know from Ref. \cite{gkp-paper} that, in well $N$,
\begin{equation}
\varphi = N\varphi_0 + \frac{\lambda_0\varphi_0}{\sqrt{2}}(a_N + a^{\dagger}_N) + \mathcal{O}(\lambda_0^2\varphi_0)\,,
\label{eqn:phi_creation_ops}
\end{equation}
where
\begin{equation}
a_N^{\dagger}\ket{N,\nu} = \sqrt{\nu+1}\ket{N,\nu+1} \qquad a_N\ket{N,\nu} = \sqrt{\nu}\ket{N,\nu-1}
\label{eqn:creation_ops}
\end{equation}
are the  effective creation and annihilation operators in well $N$. Note that they are not exactly identical to canonical bosonic creation and annihilation operator, since they are defined in terms of the  eigenstates $\{|N,\nu\}$ that contain weak nonlinear dressings relative to Harmonic oscillator eigenstates, due to the well potentials of $H_{\rm LCJ}$ being cosines, rather than  pure quadratic. Thus, we have in well $N$:
\begin{equation}
    \varphi^k \approx \varphi_0^k\left(N + \frac{\lambda_0}{\sqrt{2}}(a_N + a_N^{\dagger})\right)^k = \varphi_0^k \sum_{m=0}^k\begin{pmatrix} k \\ m\end{pmatrix} \frac{N^{k-m}\lambda_0^m}{2^{m/2}}(a_N + a_N^{\dagger})^m \,.
    \label{eqn:phik_creation_ops}
\end{equation}
For a given matrix element $\bra{N,\mu}\varphi^k\ket{N,\nu}$, with $\mu \neq \nu$, the lowest order (in $\lambda_0$) matrix element will have $|\nu-\mu|$ creation or annihilation operators (if $\mu > \nu$ or $\nu < \nu'$, respectively).  Thus, we have to leading order:
\begin{equation}
\bra{N,\mu}\varphi^k\ket{N,\nu}  \approx 
\varphi_0^{k}\begin{pmatrix} k \\ |\nu - \mu|\end{pmatrix}\frac{N^{k-|\nu-\mu|}\lambda_0^{|\nu-\mu|}}{2^{|\nu-\mu|/2}}\sqrt{\frac{\max(\mu,\nu)!}{\min(\mu,\nu)!}} 
\label{eqn:matrix_elem_leading_order}
\end{equation}
with $\begin{pmatrix} k \\ m\end{pmatrix}=0$ for $m>k$, such that the overlap above vanishes if $k<|\mu-\nu|$. 
Furthermore, from Ref. \cite{gkp-paper} we also have that
\begin{equation}
E_{N,\nu} = N^2 \varepsilon_{\text{L}} + \nu \varepsilon_0 + \mathcal{O}(\lambda_0^2)\,,
\label{eqn:well_energies}
\end{equation}
with $\varepsilon_{\text{L}} = \varphi_0^2/2L$ and $\varepsilon_0 = \sqrt{4e^2 J/C}$.
So, the left side of \eqref{eqn:pert_condition} becomes

\begin{align*}
\frac{V_k}{\varphi_0^k}\frac{\bra{N,\mu}\varphi^k\ket{N,\nu}}{E_{N,\mu} - E_{N,\nu}} &\approx V_k \begin{pmatrix} k \\ |\nu-\mu|\end{pmatrix}\frac{N^{k-|\nu-\mu|}\lambda_0^{|\nu-\mu|}}{2^{|\nu-\mu|/2}(\nu-\mu)\varepsilon_0}\sqrt{\frac{\max(\mu,\nu)!}{\min(\mu,\nu)!}}.
\end{align*}
We can bound the above, using $\max(\mu,\nu)!/{\min(\mu,\nu)}!\leq \max(\mu,\nu)^{|\mu-\nu|}$. This gives us 
\begin{equation}
\frac{V_k}{\varphi_0^k}\frac{\bra{N,\mu}\varphi^k\ket{N,\nu}}{E_{N,\mu} - E_{N,\nu}} \lesssim  \frac{V_k}{\varepsilon_0} \begin{pmatrix} k \\ |\nu-\mu|\end{pmatrix}\frac{N^{k-|\nu-\mu|}\lambda_0^{|\nu-\mu|}}{2^{|\nu-\mu|/2}(\nu-\mu)} \big{(}\max(\nu,\mu)\big{)}^{|\nu-\mu|/2}\,,
\end{equation}
For $\mu,\nu\ll 1/\lambda_0^2$,  the right hand side obtains its maximum (with $k$ fixed) when $|\nu-\mu| = 1$. 
In this case, the condition in Eq.~\eqref{eqn:pert_condition} becomes
\begin{equation}
\frac{V_kkN^{k-1}\lambda_0}{\varepsilon_0}\sqrt{\frac{\max(\nu,\mu)}{2}} \ll 1 \,.
\label{eqn:matrix_elem_minimal_nu}
\end{equation}
Isolating $V_k$, we thus obtain the condition
\begin{equation}
  V_k \ll \frac{\varepsilon_0}{k N^{k-1}\lambda_0}\sqrt{\frac{2}{\max(\mu,\nu)}}.
\end{equation}
The GKP states stabilized by our protocol for finite resistor temperature $T$ are thermal mixtures of low-lying well eigenstates, with support confined exponentially to $\nu$ less than $\nu_{\max} \sim \kb T/\varepsilon_0$. Thus, we can bound the right side above by taking $\max(\nu,\mu) < 1+\zeta \kb T/\varepsilon_0$, where $\zeta$ is an $\mathcal{O}(1)$ constant (recall we assumed $|\mu-\nu|=1$, implying either $\mu$ or $\nu$ must be nonzero). 
This leaves us with the condition
\begin{equation}
  V_k \ll \frac{\varepsilon_0}{k N^{k-1}\lambda_0}\sqrt{\frac{2}{1+\frac{k_{\rm B}T}{\zeta \varepsilon_0}}}.
\end{equation}

To obtain a meaningful bound for $V_k$ from the above, we now move on to bound $N$.
To this end, we know that the envelope of the GKP state decays in $\varphi$ over a characteristic length scale $\kappa\varphi_0$~\cite{gkp-paper}, where 
\begin{equation}
\kappa \equiv  \sqrt{\coth(2\varepsilon_0/\kb T)}/\pi\lambda_0 \,.
\end{equation}
If we demand that \eqref{eqn:pert_condition} should be satisfied up to $N \approx \alpha\kappa$---with $\alpha$ an $\mathcal{O}(1)$ constant---then this means $N \lesssim \alpha\sqrt{\coth(2\varepsilon_0/\kb T)}/\pi\lambda_0$. Using this above yields~\footnote{Note that the $m = |\nu-\mu|$ term in \eqref{eqn:phik_creation_ops} is still the leading term in $\lambda_0$ when $N \propto 1/\lambda_0$, so \eqref{eqn:matrix_elem_leading_order} and \eqref{eqn:matrix_elem_minimal_nu} still capture the leading order contribution to the matrix elements.} 
\begin{equation}
  V_k \ll \frac{\varepsilon_0 \lambda_0^{k-2}\pi^{k-1}}{\alpha^{k-1}c_T^{k-1}}\sqrt{\frac{2}{1+\frac{k_{\rm B}T}{\zeta \varepsilon_0}}}\,,
\end{equation}
where $c_T = \sqrt{\coth(2\varepsilon_0/\kb T)}$.
Finally, using $\varepsilon_0 = h\fLC/(\pi \lambda_0^2)$, we arrive at
\begin{equation}
  V_k \ll \frac{h\fLC \lambda_0^{k-4}\pi^{k-2}}{\alpha^{k-1}c_T^{k-1}}\sqrt{\frac{1}{1+{k_{\rm B}T \pi \lambda_0^2}/{\zeta h\fLC}}}.
\label{eqn:coefficient_constraint}
\end{equation}
In particular, we require 
\begin{equation}
  V_4 \ll h\fLC\frac{ \pi^{2}}{\alpha^{3}c_T^{3}}\sqrt{\frac{1}{1+{k_{\rm B}T \pi \lambda_0^2}/{\zeta h\fLC}}}
\label{eqa:cond}\end{equation}
so that the fastest possible $\sqrt{T}$ gate time will be expected to scale with circuit parameters as 
\begin{equation}
t_{4,{\rm min}} \sim \frac{C}{\fLC}\sqrt{1+\frac{k_{\rm B}T \pi \lambda_0^2}{\zeta h\fLC}}'
\label{eqa:t4min}
\end{equation}
where we used $c_T \geq 1$, and $C =\alpha^3/\pi^2$ is some prefactor of magnitude $\gtrsim \mathcal O(1)$. 
In particular, at low temperatures, $k_{\rm B}T\ll h\fLC/\lambda_0^2$, we have $t_{4,{\rm min}}\sim 1/\fLC$. 

\subsection{Dephasing due to residual non-quartic coefficients}
\label{seca:phases}
In addition to modifying the eigenstates, the residual non-quartic terms in the effective potential can also decohere the qubit by imparting different phases on states in different wells. If the $V_k$ for $k \neq 4$ are small enough,  these phases will not take the system out of the codespace, and are correctable via dissipation. We now seek to identify conditions that will ensure this is the case.
First, let us consider a generic $\ket{\!+\!z}$ GKP state with wavefunction $\psi_{+z}(\varphi)$: 
\begin{equation}
\psi_{+z}(\varphi) = \mathcal{N}\chi(\varphi)\sum_N \eta(\varphi - 2N\varphi_0)\,,
\label{eqn:psi_GKP_even}
\end{equation}
where $\mathcal{N}$ is a normalization constant, $\eta(\varphi)$ is the system wavefunction within each well, and $\chi(\varphi)$ the envelope. To leading perturbative order, the system wavefunction in well $N$ acquires a dynamical phase $\alpha_k N^k$, where $\alpha_k := V_k\,t/\hbar$, from evolution for a time $t$ under the perturbation $V_k (\varphi/\varphi_0)^k$. Assuming the system wavefunction is otherwise unaffected (which will be the case if $\eta$ is an eigenstate of of the system Hamiltonian and $t$ a multiple of the revival time), the wavefunction for the system is:
\begin{equation}
    \psi_{+z}(\varphi,t) = \mathcal{N}\chi(\varphi)\sum_N e^{-i\alpha_k N^k} \eta(\varphi - 2N\varphi_0) \,.
    \label{eqn:psi_GKP_with_evolved}
\end{equation}
We want to ensure that the spurious phases do not take the state of the system out of the codespace. At this level of approximation, the support of the state over the spectrum of $S_1$ is unaffected; the phase factors $\exp(-i \alpha_kN^k)$ are \textit{a global phase within each well}, and thus cannot delocalize the system within each well in flux. It remains to check whether the support of the system over the spectrum of $S_2$---i.e., the support of the system within each well in charge space---is affected.   

To check this, we need to examine the Fourier transform of the state. To facilitate this calculation, we make the additional approximation
\begin{equation}
    \psi_{+z}(\varphi,t) \approx \mathcal{N}\chi(\varphi) e^{-i \alpha_k(\varphi/\varphi_0)^k}\sum_N\eta(\varphi-2N\varphi_0)\,.
    \label{eqn:psi_GKP_with_evolved_approx}
\end{equation}
This approximation is justified when $\alpha_k (\varphi/\varphi_0)^k$ changes little over the range of support of $\eta$ in each well; we will come back to this point later. 
With this approximation, let us compute the Fourier transform:
\begin{equation}
    \mathcal{F}[\psi_{+z}](q) = e\mathcal{N}\sum_M \mathcal{F}[\tilde{\chi}](q - Me)\,\mathcal{F}[\eta]\left(\frac{Me}{\hbar}\right)\,,
    \label{eqn:psi_GKP_FT}
\end{equation}
where 
\begin{equation}
\mathcal{F}[f](q) = \frac{1}{\sqrt{2\pi}}\int_{-\infty}^{\infty} e^{-iq\varphi/\hbar}f(\varphi)\,\mathrm{d}\varphi
\label{eqn:FT}
\end{equation}
denotes the Fourier transform of a function $f$, and $\tilde{\chi}(\varphi) = \chi(\varphi)e^{-i\alpha_k (\varphi/\varphi_0)^k}$. In deriving \eqref{eqn:psi_GKP_FT}, we have used the convolution theorem, and the fact that the Fourier transform of $\sum_N \eta(\varphi-2N\varphi_0)$ is the product of $\mathcal{F}[\eta]$ and a Dirac comb of periodicity $\pi\hbar/\varphi_0 = e$. Evidently, the Fourier transform of $\tilde{\chi}$ describes the localization of $ \mathcal{F}[\psi_{+z}]$ in each well in charge space, which is exactly what we wish to probe.

\begin{figure}
    \centering
    \includegraphics[scale=0.66]{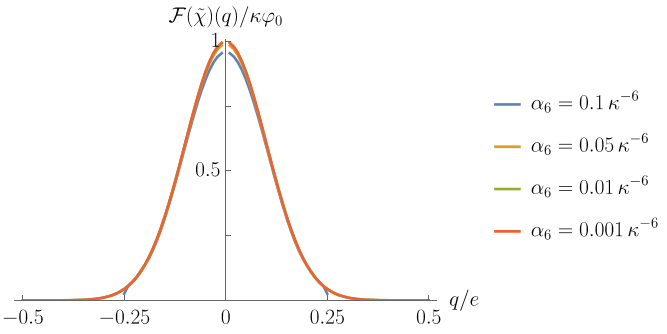}
    \caption{Fourier transform $\mathcal{F}(\tilde{\chi})$ of the modified envelope $\tilde{\chi}$ [given by \eqref{eqn:chi_tilde_FT}] for $k=6$ and $\kappa = 10/\pi$ as a function of $q$, for different values of $\alpha_6$. We see rapid convergence of these functions as $\alpha_6\kappa^6$ becomes small, in line with the constraint \eqref{eqn:coeff_condition_FT}.}
    \label{fig:well_wvfn_renorm_series}
\end{figure}

To obtain some concrete results, let us now take $\chi$ to be a Gaussian with width $\kappa\varphi_0$: $\chi(\varphi) = \exp[-\varphi^2/(2\kappa^2\varphi_0^2)]$. We then compute $\mathcal{F}(\tilde{\chi})$ by a Taylor expansion of $\exp(-i\alpha_k(\varphi/\varphi_0)^k)$:
\begin{align}
    \mathcal{F}[\tilde{\chi}](q) &= \frac{1}{\sqrt{2\pi}}\sum_m \frac{(-i \alpha_k/\varphi_0^k)^m}{m!}\int_{-\infty}^{\infty}\varphi^{km}e^{-\varphi^2/(2\kappa^2\varphi_0^2)}e^{-iq\varphi/\hbar}\,\mathrm{d}\varphi \notag\\
    &= \kappa\varphi_0 e^{-(\pi\kappa q)^2/2e^2}\sum_m\frac{1}{m!}\left(-i\alpha_k \kappa^k\right)^m\sum_{n=0}^{\lfloor mk/2\rfloor}\frac{(mk)!}{n!(mk-n)!2^n}\left(-i\frac{\pi \kappa q}{e}\right)^{mk-2n} \,.\label{eqn:chi_tilde_FT}
\end{align}
In deriving this result, we have used general results for Gaussian integrals \cite{Wolfram-Gaussian-Integral}. We see that $\mathcal{F}[\tilde{\chi}]$ is the product of a Gaussian with decay width $\hbar/\kappa\varphi_0 = e/\kappa\pi$, and the function of $q$ captured by the summations. The function of $q$ in the summations will, generically, renormalize the width of the wavefunction in each well in charge space. If the width exceeds $\sim e/2$, the system leaves the code subspace, and the encoded information is lost. 
However, by inspecting \eqref{eqn:chi_tilde_FT}, we see that these corrections will be suppressed if
\begin{equation}
    \alpha_k \kappa^k = \frac{V_k \kappa^k t}{\hbar} \ll 1\,.
    \label{eqn:coeff_condition_FT}
\end{equation}
We illustrate this by plotting $\mathcal{F}[\tilde{\chi}](q)$ for several $\alpha_k$ in Fig. \ref{fig:well_wvfn_renorm_series}. Intuitively, the condition \eqref{eqn:coeff_condition_FT} requires that the dynamical phase from the perturbation $V_k \varphi^k$ after time $t$ is small in all wells that the system has support on; this range of support is characterized by the scale of the envelope $\sim \kappa \varphi_0$. 
Hence, for $k\neq 2, 4$, we require 
\begin{equation}
V_k  \ll \frac{\hbar}{\kappa^k t_{4}}\,,
\label{eqa:cond2}
\end{equation}
where 
\begin{equation}\kappa \equiv \sqrt{\coth(2\varepsilon_0/\kb T)}/\pi\lambda_0
\end{equation}
denotes the envelope width of the GKP state in units of $\varphi_0$. 
Together with Eq.~\eqref{eqa:cond} this consitutes the conditions on the non-quartic potential of the nonlinear potential. 
Since $t_4  $ can be chosen to be of order $1/\fLC$,  the condition in Eq.~\eqref{eqa:cond2} is generally stricter. 
\subsection{Justification of  Eq.~\eqref{eqn:psi_GKP_with_evolved_approx}}
Here we provide justification for the approximation made in \eqref{eqn:psi_GKP_with_evolved_approx}. The replacement $e^{-i\alpha_k N^k} \rightarrow e^{-i\alpha_k (\varphi/\varphi_0)^k}$ is justified when $|e^{-i\alpha_k N^k} - e^{-i\alpha_k(\varphi/\varphi_0)^k}| \ll 1$ over all $\varphi$ the system wavefunction has support on. To see that this is the case, we can expand $e^{-i\alpha_k(\varphi/\varphi_0)^k}$ about $\varphi = N\varphi_0$:
\begin{equation}
    e^{-i\alpha_k N^k} - e^{-i\alpha_k(\varphi/\varphi_0)^k} = e^{-i\alpha_k N^k}\sum_{m=1}^{\infty} \frac{1}{m!}\left(\frac{(-i\alpha_k)^{1/k}}{\varphi_0}\right)^m \mathcal{P}^k_m((-i\alpha_k)^{1/k}N) (\varphi - N\varphi_0)^m
    \label{eqn:exp_series}
\end{equation}
where 
\begin{equation}
    \mathcal{P}^k_m(x) = e^{-x^k}\frac{\mathrm d^m}{\mathrm dx^m} e^{x^k}\,.
    \label{eqn:exp_deriv}
\end{equation}
Since $\mathcal{P}^k_m$ is a polynomial of finite degree, $(\alpha_k)^{1/k}N \ll 1$ (by virtue of $N \lesssim \kappa$ and \eqref{eqn:coeff_condition_FT}), $(\alpha_k)^{1/k} \ll 1$ (from \eqref{eqn:coeff_condition_FT}, assuming $\kappa \sim \mathcal{O}(1)$), and $(\varphi - N\varphi_0)^m/\varphi_0^m \lesssim \lambda_0 \ll 1$, the difference \eqref{eqn:exp_series} will be small. The replacement $e^{-i\alpha_k N^k} \rightarrow e^{-i\alpha_k (\varphi/\varphi_0)^k}$ is thus justified, provided \eqref{eqn:coeff_condition_FT} holds.  

\section{Simulation of flux noise}
We incorporate flux noise in our simulations via a time-dependent fluctuating term in the Hamiltonian
\begin{equation}
        H_{\rm noise} = \frac{\varphi}{L} \xi_{\varphi}(t)\,,
        \label{eq:noise_Ham}
\end{equation}
    where $\xi_{\varphi}$ is a zero-mean fluctuating scalar signal with autocorrelation
\begin{equation}
    \avg{\xi_{\varphi}(t')\xi_{\varphi}(t)} = \gamma_{\varphi} J(t'-t) = \frac{\gamma_{\varphi}}{2\pi} \int_{-\infty}^{\infty} \tilde{J}(\omega)e^{-i\omega(t-t')}\,{\rm d}\omega
    ,.\label{eqn:autocorrelation_phi}
\end{equation}
The parameter $\gamma_{\varphi}$ is the noise strength, and characterizes the strength of the power spectral density of the autocorrelation.

In experiments with superconducting circuits, the observed noise spectral density is dominated by low-frequency components, and has a $1/f$ structure \cite{quant-engineers-guide}. To model this, we take our noise to have the following spectral density:
\begin{equation}
    \tilde{J}(\omega) = \frac{\Omega}{|\omega| + \omega_0} e^{-|\omega|/\Lambda}\,,
\label{eqn:spectral_fn}
\end{equation}
where $\Omega$ is a frequency scale (inserted for convenience), and $\omega_0$, $\Lambda$ are low and high-frequency (IR and UV) cutoffs, respectively. This spectral function matches noise of the $1/f$ type observed in experiments \cite{quant-engineers-guide}, down to the IR cutoff $\omega_0$ and up to the UV cutoff $\Lambda$. These cutoffs are needed to ensure convergence of the integral in \eqref{eqn:autocorrelation_phi} and a finite value of $J(0)$, respectively. Ultimately, we will take $\Omega = 2\pi\,{\rm Hz}$ (so that $\gamma_{\varphi}$ measures the power spectral density at frequency $1$ Hz), and $\omega_0 = 10^{-4}\,{\rm Hz}$, $\Lambda = 1$ MHz.\footnote{Experiments show the $1/f$ form of the spectral function to hold to frequencies as low as $\sim$ milliHertz, and up to qubit frequencies ($\sim$ MHz-GHz) \cite{quant-engineers-guide}, hence the choice of these cutoffs.} 

\subsection{Generating $\xi(t)$}
Here, we detail how we compute the noise signal $\xi_{\varphi}(t)$.

Due to the long-lived correlations of $\xi(t)$, we must generate the signal for the entirety of the simulation duration.
To do this, we generate a white noise signal $\chi(t)$ and convolve it with the \textit{jump correlator} $g(t)$ (borrowing the language of Ref. \cite{ULE_paper})---i.e., we compute
\begin{equation}
    \xi(t) = \int_{-\infty}^{\infty} g(t-\tau) \chi(\tau)\, \mathrm{d}\tau \,,
\label{eqn:xi_convolution}
\end{equation}
where
\begin{equation}
    g(t) = \frac{1}{\sqrt{2\pi}}\int_{-\infty}^{\infty}\sqrt{\tilde{J}(\omega)}\, e^{-i \omega t}\,\mathrm{d}\omega\,.
    \label{eqn:jump_corr}
\end{equation}
This way, the autocorrelation is:
\begin{align}
    \avg{\xi(t')\xi(t)} &= \int_{-\infty}^{\infty}\int_{-\infty}^{\infty} g(t'-\tau')g(t-\tau) \underbrace{\avg{\chi(\tau')\chi(\tau)}}_{\delta(\tau'-\tau)} \,\mathrm{d}\tau'\,\mathrm{d}\tau \notag\\
    &= \int_{-\infty}^{\infty} g(t'-\tau)g(t-\tau)\,\mathrm{d}\tau \notag\\
    &= J(t-t')\,,
\end{align}
where the last equality can be derived by inserting \eqref{eqn:jump_corr} and using the fact that $\tilde{J}(-\omega) = \tilde{J}(\omega)$. So we see that \eqref{eqn:xi_convolution} generates the desired signal, provided $\avg{\chi(t)} = 0$ for all $t$.

Numerically, we must discretize the integral in \eqref{eqn:xi_convolution} and compute the convolution numerically:
\begin{equation}
    \xi(t) = \sum_m g(t-t_m) \chi_m \sqrt{\Delta t_m}
    \label{eqn:xi_numerical}
\end{equation}
where $\Delta t_m = t_{m+1}-t_m$, and the $\{\chi_m\}_m$ are i.i.d. with zero mean and unit variance.
To compute $\xi(t)$ via \eqref{eqn:xi_numerical}, we need the jump correlator $g(t)$. It is always possible to discretize the integral in \eqref{eqn:jump_corr} and compute $g(t)$ numerically from $\tilde{J}$, but for the spectral function \eqref{eqn:spectral_fn} we can compute $g(t)$ analytically:
\begin{align}
    g(t) &= \frac{1}{\sqrt{2\pi}}\int_{-\infty}^{\infty} \frac{\sqrt{\Omega}}{\sqrt{|\omega|+\omega_0}} e^{-i \omega t} e^{-|\omega|/2\Lambda} \mathrm{d}\omega \notag\\
     &= \sqrt{2\Omega}\, e^{\omega_0/2\Lambda}\,\text{Re}\left[\frac{e^{i\omega_0 t}}{\sqrt{it+1/2\Lambda}}\,\text{Erfc}(\sqrt{it+1/2\Lambda}\sqrt{\omega_0})\right] \label{eqn:jump_corr_analytic}\,, 
\end{align}
where 
$$\text{Erfc}(z) = 1 - \text{Erf}(z) = \frac{2}{\sqrt{\pi}}\int_z^{\infty} e^{-y^2}\,{\rm d}y$$
is the complimentary error function. Using the analytic formula \eqref{eqn:jump_corr_analytic} in \eqref{eqn:xi_numerical}, we can efficiently compute different realizations of the noise signals $\xi_{\varphi}(t)$ by sampling the $\{\chi_m\}_m$. 

For our simulations, we pick the grid of points $\{t_m\}_m$ to be the union of two subgrids: the first subgrid consists of evenly spaced $t_m$ between 0 and 1 second with resolution $1/2\Lambda$, and the second subgrid consists of evenly spaced $t_m$ from 1 second to $1/\omega_0$, with resolution $0.4$ seconds. For each realization of the noise signal, the white noise variables $\chi_m$ are sampled from a standard normal distribution. We then evaluate \eqref{eqn:xi_numerical} for 101 times $t$ uniformly spaced between 0 and $10^{-4}$ seconds. To access $\xi(t)$ at different times during the simulation, we create and evaluate an interpolating function.  

\subsection{Including $\xi_{\varphi}(t)$ in simulations}
Now having access to realizations of the noise signals $\xi_{\varphi}(t)$, we outline how the noise is included in the simulations of the $\sqrt{T}$ gate protocol.

\subsubsection{Stabilizer and $\varphi^4$ Segments}
During the stabilizer and $\varphi^4$ segments, we have the following system Hamiltonian:
\begin{equation}
    H =  \HLCJ + V(\varphi) + \frac{\xi_{\varphi}(t)}{L}\varphi 
    \label{eqn:H_sys_noise}
\end{equation}
where $V(\varphi)$ is the (effective) potential in $\varphi$  arising from the ancillary Josephson junctions (if any). 

This Hamiltonian, for a given realization of $\xi_{\varphi}$ along with the associated jump operators $\{L_i(t)\}_i$, defines a master equation with a time-dependent Liouvillian.
In principle, we can proceed with wavefunction evolution using the SSE formalism with a time-dependent effective Hamiltonian $H_{\text{eff}}(t)$. 
However, since the system Hamiltonian $H$ is time-dependent, but not periodic, in general one must explicitly integrate the resulting differential equation to simulate time evolution. Since the $\varphi^4$ segment is typically long ($\sim 10^3-10^4$ times longer than the stabilizer segments in the native stabilization protocol), exact evolution in this way is \textit{extremely} costly. We can achieve a dramatic speedup by doing the approximation below.
\begin{enumerate}
    \item Simulate SSE evolution from $t_1$ to $t_2$ in the absence of noise, i.e. with $\chi_{\varphi}(t) \equiv 0$.
    \item Act on the state with a unitary $\exp(-i \alpha(t_1,t_2) \varphi)$, where
    \begin{equation}
    \alpha(t_1,t_2) = \frac{1}{\hbar L}\int_{t_1}^{t_2} \xi_{\varphi}(t)\,\mathrm{d}t\,.
    \label{eqn:alpha}
    \end{equation}
\end{enumerate}
Step 1 can be done efficiently, by pre-computing and storing logarithmically spaced evolution operators, and step 2 involves only a single integral of a scalar function (rather than an integral of a large system of differential equations). This procedure takes orders of magnitude less time than the exact integration of the differential equation generated by $H_{\text{eff}}(t)$.

This method is not exact. To understand the nature of the error incurred by this approximation, let us go to the co-rotating frame of the noise Hamiltonian $\xi_{\varphi}(t)\varphi/L$. In this frame, the system Hamiltonian is
\begin{equation}
    \tilde{H}(t) = \HLCJ - \frac{q}{C}\hbar\alpha(t_1,t) + \frac{\hbar^2\alpha^2(t_1,t)}{2C}\mathbb{I}\,,
    \label{eqn:H_rot_noise}
\end{equation}
and the jump operator from the resistor is \cite{ULE_paper}
\begin{equation}
    \tilde{L}(t) \propto \int_{-\infty}^{\infty} g(t-s) \tilde{U}(t,s)[q - \hbar\alpha(t_1,s)\mathbb{I}]\tilde{U}(t,s)^{\dagger}\,\mathrm{d}s\,,
    \label{eqn:L_rot_noise}
\end{equation}
where $\tilde{U}(t,s) = \mathcal{T}\{\exp(-\frac{i}{\hbar}\int_s^t \tilde{H}(\tau)\,\mathrm{d}\tau)\}$ is the time evolution operator in the rotating frame.
Evolving the system according to steps 1 and 2 above corresponds to discarding the terms proportional to $\alpha$ above, which are a random drift of the charge operator due to the noise signal. Heuristically, this random drift of the charge operator will induce  diffusion of the state along $\varphi$, which we expect the resistor can still correct if the noise signal is sufficiently small.

\subsubsection{Free Segment}
During the free segment, we model the system as evolving (\textit{unitarily}) under the Hamiltonian
\begin{equation}
    H = \frac {\varphi^2}{2L} + \frac{q^2}{2C} +  \frac{\varphi}{L}\xi_{\varphi}(t)\,.
\label{eqn:H_free}
\end{equation}
In the co-rotating frame of the LC Hamiltonian $H_{\text{LC}}$, we have:
\begin{equation}
    \tilde{H}(t) = \frac{1}{L}\left[\cos(2\pi \fLC t)\, \varphi + \frac{\varphi_0 \nu}{2e}\sin(2\pi \fLC t)\,q\right] \xi_{\varphi}(t)
    \label{eqn:H_rot}
\end{equation}
where $\displaystyle \nu = \frac{4e^2}{h}\sqrt{\frac{L}{C}}$ is the oscillator impedance is units of the quantum resistance for Cooper pairs. Time evolution under this Hamiltonian yields (up to a time-dependent global phase)
\begin{align}
    \tilde{U}(t) &= \mathcal{T} \exp\left(-\frac{i}{\hbar}\int_0^t \tilde{H}(t')\,\mathrm{d}t'\right) \notag\\
    &= \exp\bigg{\{}-\frac{i}{\hbar}\bigg{[}\int_0^t\left(\frac{\varphi_0 \nu}{2eL}\sin(2\pi\fLC t') \xi_{\varphi}(t')\right) \mathrm{d}t'\, q + \int_0^t\left(\frac{1}{L}\cos(2\pi\fLC t')\xi_{\varphi}(t')\right) \mathrm{d}t'\, \varphi \bigg{]}\bigg{\}}\,, \label{eqn:rot_frame_unitary}
\end{align}
where the second equality follows from the first by virtue of the fact that $[q,\varphi] = i\hbar\mathbb{I}$ is a multiple of the identity. We thus have a random unitary of the form $\exp(-i (A(t)q + B(t)\varphi)/\hbar)$, parameterized by random variables
\begin{subequations}
    \begin{align}
        A(t) &= \frac{\varphi_0 \nu}{2eL}\int_0^t\sin(2\pi\fLC t') \xi_{\varphi}(t') \mathrm{d}t' \\
        B(t) &= \frac{1}{L}\int_0^t \cos(2\pi\fLC t')\xi_{\varphi}(t') \mathrm{d}t'\,.
    \end{align}
\end{subequations}
Since we have access to the noise signal $\xi_{\varphi}(t)$, we can compute the coefficients $A(t)$ and $B(t)$ via explicit integration, and construct $\tilde{U}(t)$. The full evolution during the free segment is then obtained by going back to the lab frame:
\begin{equation}
    U(t) = \exp\left(-\frac{i}{\hbar} H_{\text{LC}} t\right) \tilde{U}(t) = \exp\left(-\frac{i}{\hbar} H_{\text{LC}} t\right) \exp\left(-\frac{i}{\hbar}\left[A(t)q + B(t)\varphi\right]\right)\,.
\label{free_segment_evo}
\end{equation}
We emphasize that this method is \textit{exact}; unlike in the stabilizer and $\varphi^4$ segments, there is no error incurred by simulating in this manner.

\bibliography{bibliography_T_gate.bib}{}
\bibliographystyle{ieeetr}